\documentclass[reprint,aps,pra,floatfix]{revtex4-1}
\usepackage{graphicx}
\usepackage{dcolumn}
\usepackage{bm}
\usepackage{hyperref}
\usepackage{silence}
\usepackage{float}
\usepackage{booktabs}
\usepackage{textgreek}
\usepackage{threeparttable}
\usepackage{amsmath}
\usepackage{xcolor}

\hypersetup{colorlinks, citecolor=blue, linkcolor=blue, urlcolor=blue}

\usepackage{graphicx}

\begin{document}
\title{Bound Dark Energy: a particle's origin of dark energy }

\author{Axel de la Macorra}%
\email{macorra@fisica.unam.mx}
\author{Jose Agustin Lozano Torres}
 \email{jalozanotorres@gmail.com}

\affiliation{ Instituto de Física, Universidad Nacional Autónoma de México, Ciudad de México, 04510, México}




\date{\today}

\begin{abstract}
 \noindent Dark energy, the enigmatic force driving the accelerated cosmic expansion of the universe, is conventionally described as a cosmological constant in the standard $\Lambda$CDM model. However, measurements from the Dark Energy Spectroscopic Instrument (DESI) reports a $>2.5\sigma$ preference for dynamical dark energy, with baryon acoustic oscillation (BAO) data favoring a time-varying equation of state $w(z)$ over the cosmological constant ($w = -1$). We present the Bound Dark Energy (BDE) model, where dark energy originates from the lightest meson field $\phi$ in a dark SU(3) gauge sector, emerging dynamically via non-perturbative interactions. Governed by an inverse-power-law potential $V(\phi) = \Lambda_c^{4+2/3}\phi^{-2/3}$, BDE has no free parameters—one less than $\Lambda$CDM and three less than $w_0w_a$CDM models. Combining the DESI BAO measurements, cosmic microwave background data, and Dark Energy Survey SN Ia distance measurements from the fifth year, BDE achieves a $42\%$ and $37\%$ reduction in the reduced $\chi^2_{\rm BAO}$ compared to $w_0w_a$CDM and $\Lambda$CDM, respectively, while having an equivalent fit for type Ia supernovae and the cosmic microwave background data. The model predicts a dark energy equation of state transitioning from radiation-like ($w = 1/3$) at early times ($a < a_c$) to $w_0 = -0.9301 \pm 0.0004$ at present time. The ($w_0,w_a$) contour is 10,000 times smaller in BDE than in $w_0w_a$CDM model, while having an equivalent cosmological fit. Key parameters - the condensation energy scale $\Lambda_c = 43.806 \pm 0.19$ eV and epoch $a_c = 2.4972 \pm 0.011 \times 10^{-6}$ - align with high-energy physics predictions. These results, consistent with current observational bounds, establish BDE as a predictive framework that unifies particle physics and cosmology, offering a first-principles resolution to dark energy’s dynamical nature.

\end{abstract}

\maketitle


\begin{figure*}[ht]
  \centering
  \begin{tabular}{c@{\hspace{1em}}c@{\hspace{1em}}c@{\hspace{1em}}c@{\hspace{1em}}c@{\hspace{1em}}c}
      \includegraphics[width=18em,height=13.7em]{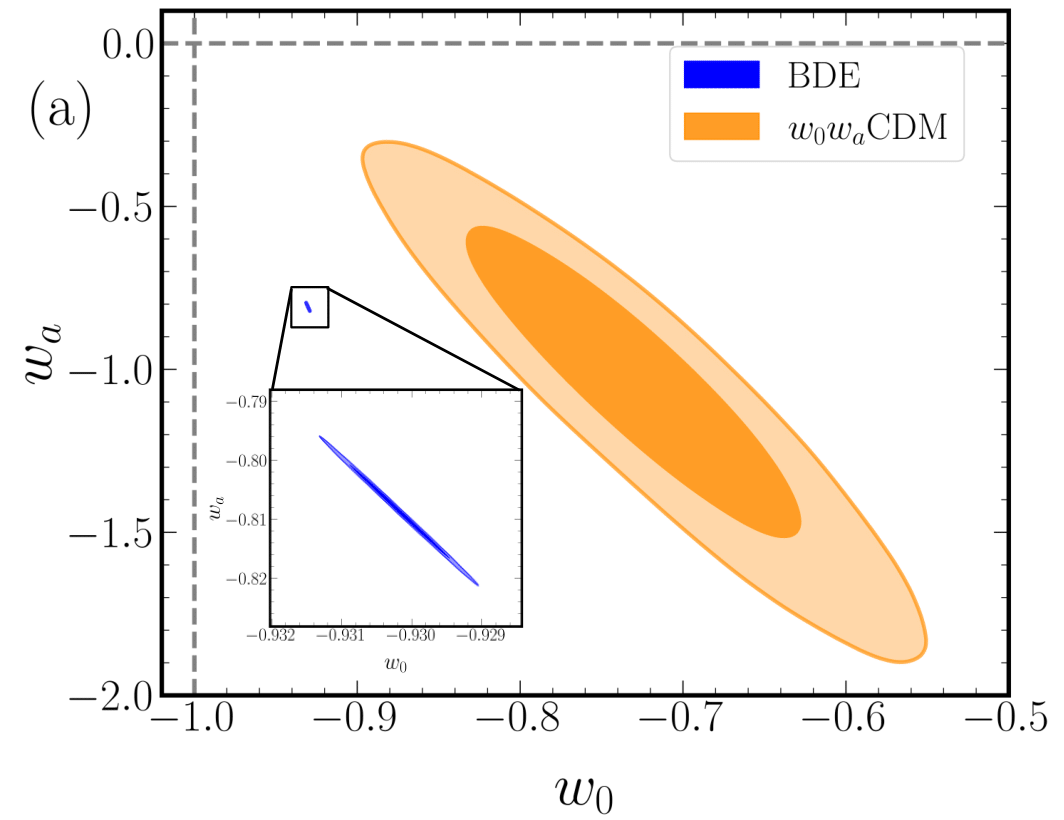} &
      \includegraphics[width=18em]{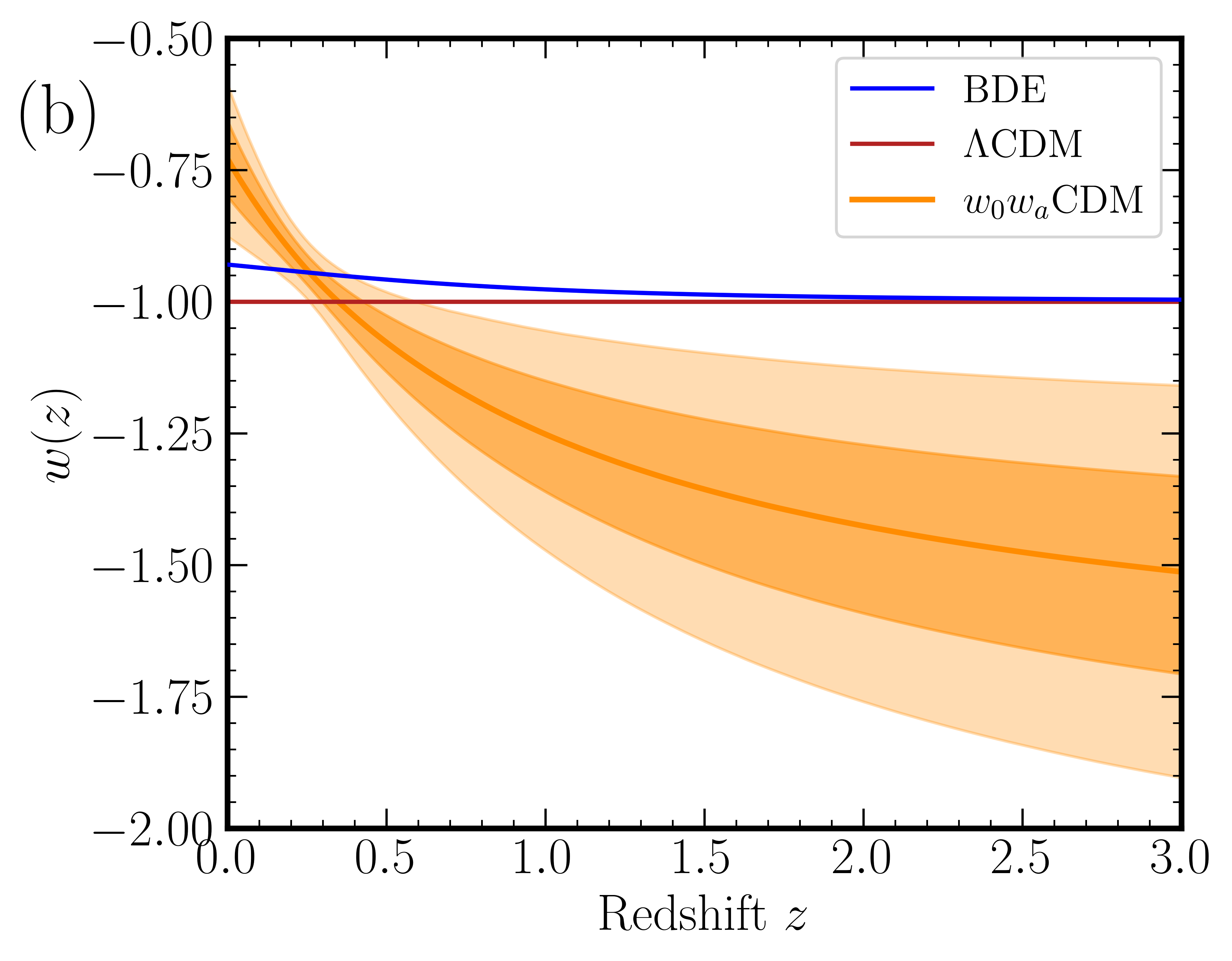} &
      \includegraphics[width=18em]{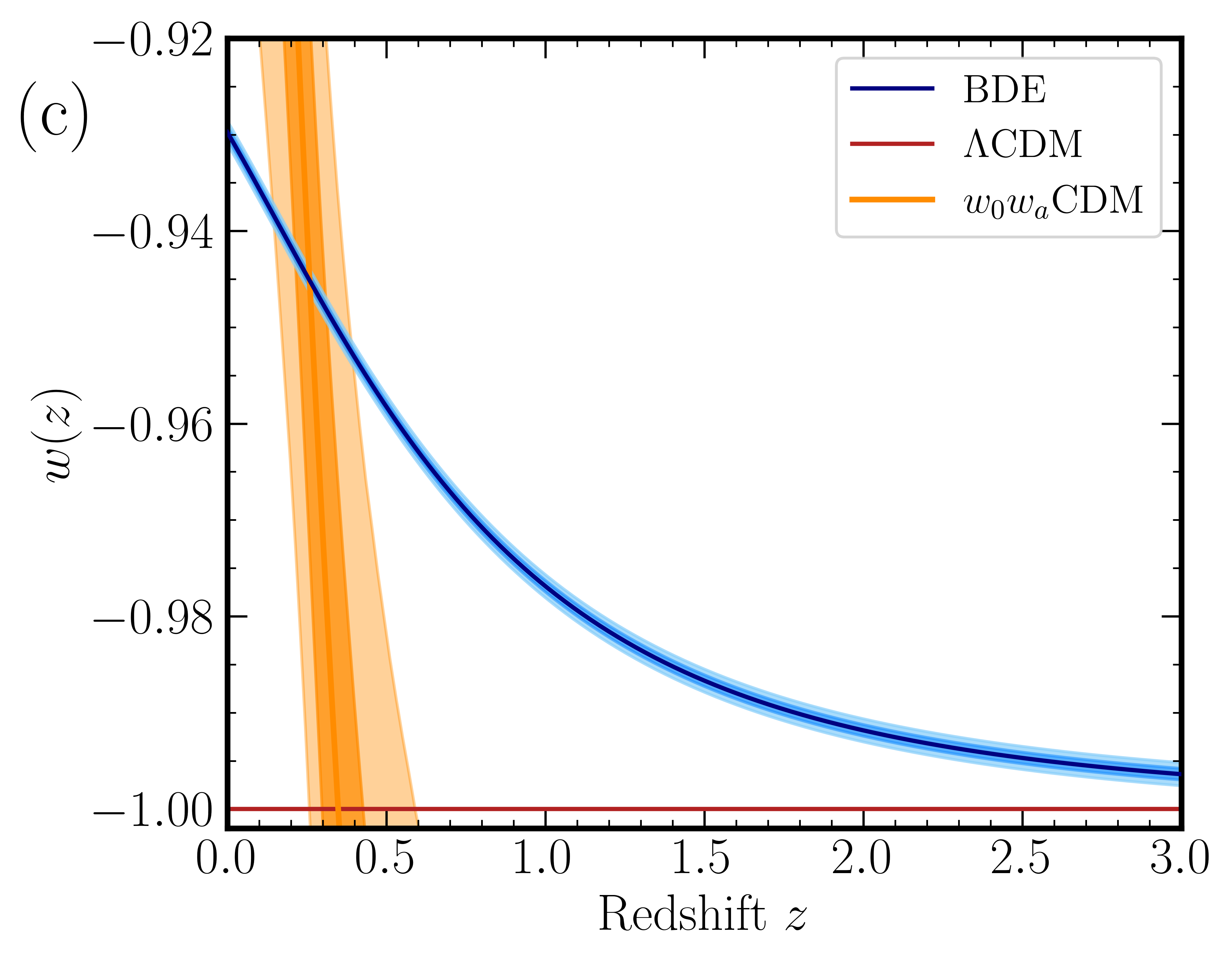}&\\
       \includegraphics[width=18em]{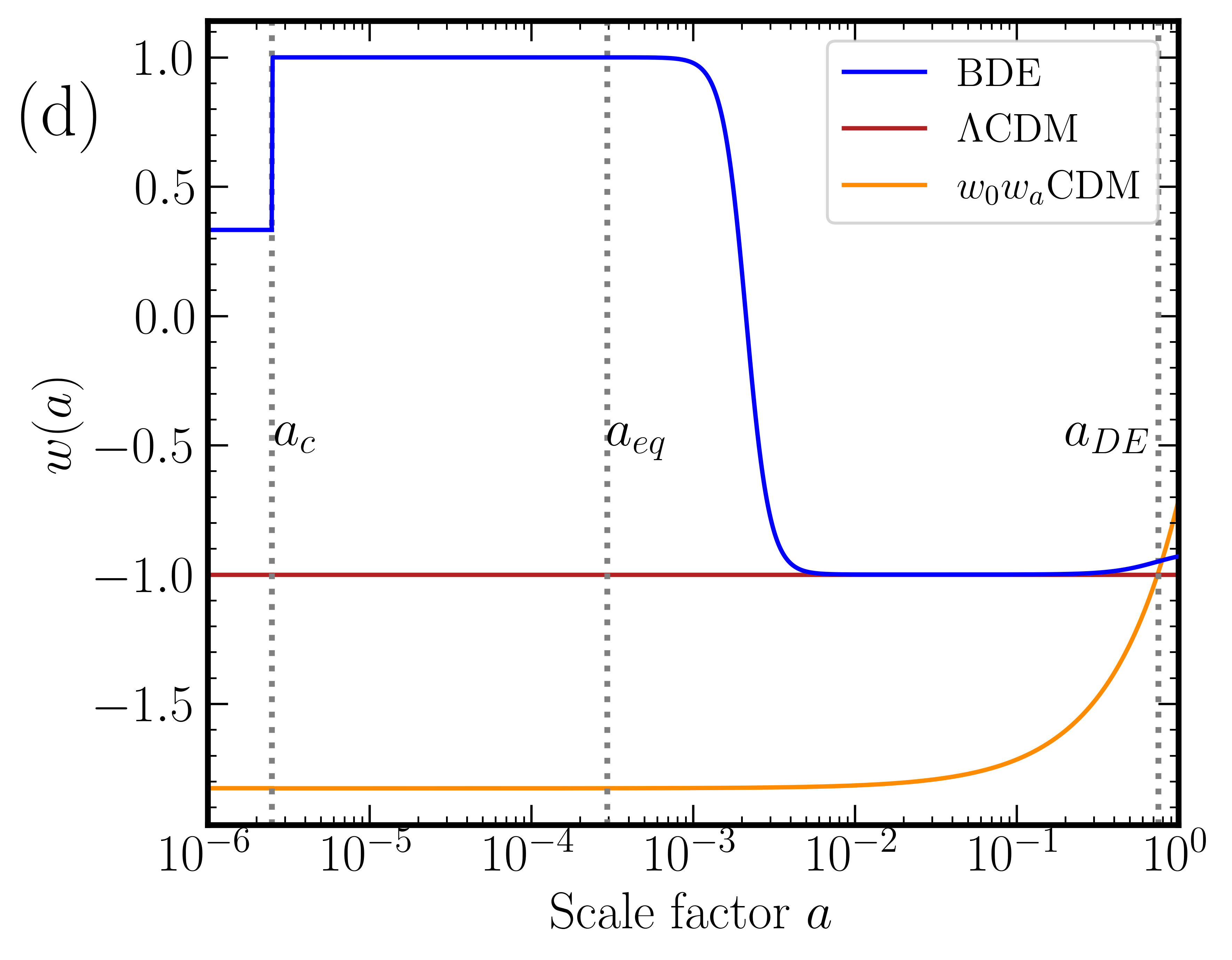} &
      \includegraphics[width=17em,height=14em]{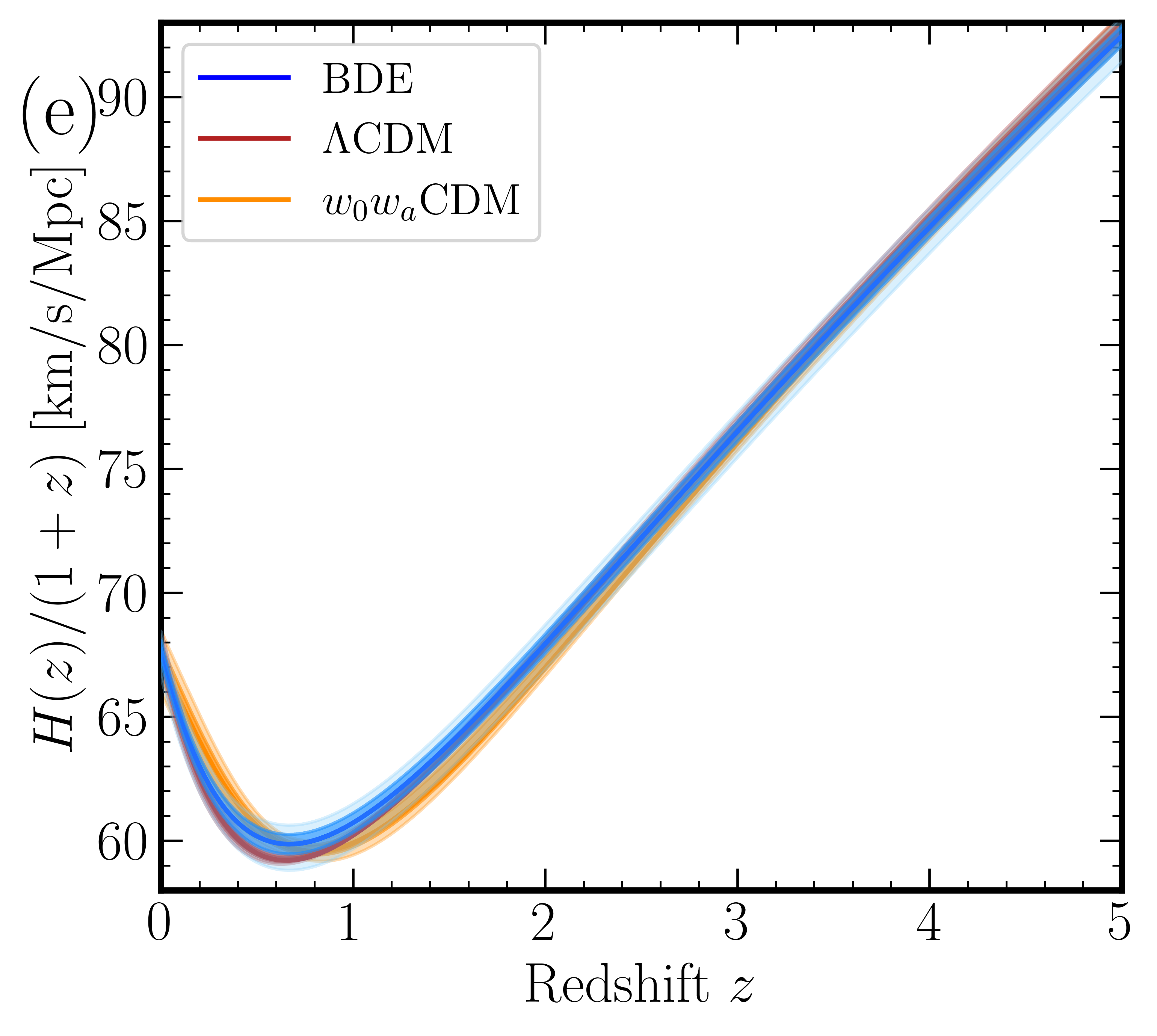} &
      \includegraphics[width=17em,height=14em]{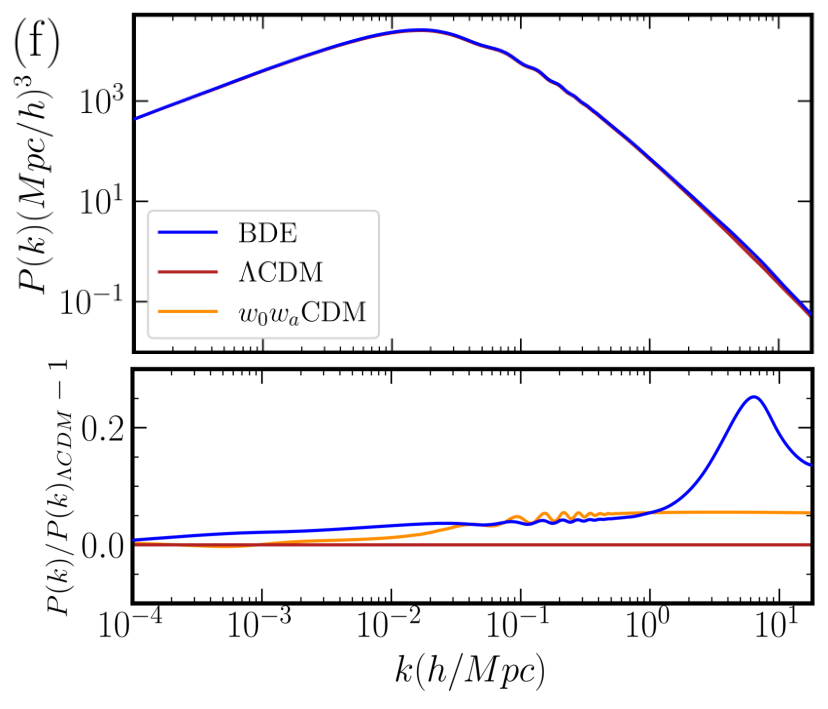} &
  \end{tabular}
  \vspace*{8pt}
  \caption{\label{Fig:1}\textbf{(a)}: Marginalized posterior constraints of $68\%$ and $95\%$ confidence levels in the $w_0-w_a$ plane for the $w_0w_a$CDM and bound dark energy models, from DESI BAO  combined with CMB and DES-SN5YR datasets. This combination favors $w_0 > -1$, $w_a < 0$, exhibiting a clear discrepancy with $\Lambda$CDM at the $\geq 2\sigma$ level. \textbf{(b)}: Comparison of the constraints on the equation of state of dark energy using the $w_0w_a$CDM model (in dark-orange) and the bound dark energy model (in blue). The contours around the best-fit from each model represent the $68\%$ and $95\%$ confidence intervals. The $\Lambda$CDM limit corresponds to the horizontal solid dark-red line. \textbf{(c)}: A closer view comparison on the bound dark energy and $w_0w_a$CDM models equation of state. \textbf{(d)}: Evolution of the equation of state of dark energy from early to present time for the given cosmological models. \textbf{(e)}: Comoving expansion rate in BDE, $\Lambda$CDM and $w_0w_a$CDM models at late times. The solid lines correspond to the best-fit model, while the contours around them represent the $68\%$ and $95\%$ confidence intervals. \textbf{(f)}: Matter power spectra at redshift $z=0$ for the best-fit across the bound dark energy, $\Lambda$CDM, and $w_0w_a$CDM models. The bottom panel shows the ratio with respect to $\Lambda$CDM of the matter power spectrum.     }
\end{figure*}

\noindent \textbf{Introduction--} The origin of the accelerated expansion of the universe has been an intriguing enigma in over two decades. Until recently the established proposal to explain this cosmic evolution was a cosmological constant.  Interesting the recent  Dark Energy Spectroscopic Instrument (DESI) precise measurements of baryon acoustic oscillations (BAO) across various cosmic tracers -such as galaxies, quasars, and the Lyman-$\alpha$ forest- reveal a preference for a time-varying equation of state parameter $w$, deviating from the conventional cosmological constant value ($w \equiv -1$) at a significance level exceeding $2.5\sigma$. This finding is situated within the $(w_0,w_a)$ parameter space - where $w_0$ and $w_a\equiv -dw/da|_{a_{o}}$ are the equation of state parameter and its first derivative at present time - favoring a dynamical dark energy behavior \cite{desicollaboration2024desi2024vicosmological}. 
Our dark energy particle referred as "Bound Dark Energy" (BDE) corresponds to the lightest meson field (denoted by $\phi$),  residing within a dark SU(3) gauge group  \cite{PhysRevD.72.043508,PhysRevLett.121.161303}. Interesting, our BDE model has no
free parameters,  one less than  $\Lambda$CDM model and three less than $w_0w_a$CDM. 
Our Bound Dark Energy model is described by the same formalism than the standard model gauge groups,  which encompasses the gauge interactions described by SU(3)$\times$SU(2)$\times$U(1) corresponding to the strong, weak and electromagnetic forces. 
The null-free parameter nature of BDE contributes to its enhanced performance, as evidenced by the remarkable size of the confidence contours areas in the $w_0 - w_a$ plane which is approximately 10,000 times smaller in BDE than in the $w_0 w_a$CDM model as shown in Fig.\ref{Fig:1}\textbf{(a)}.This considerable decrease in contour area underscores vital attributes of the BDE model: its robustness and absence of free parameters compared  $w_0w_a$CDM model and $\Lambda$CDM.    

\noindent The excellent cosmological fit of BDE is confirmed by lowering the reduced $\chi^2_{BAO}$ by  42\% compared to $w_ow_a$CDM having three free parameters less and by diminishing by 37\%  with respect to $\Lambda$CDM with one free parameter less, as detailed in the \textit{Results} section. By utilizing the exceptional first-year data release from DESI, along with complementary cosmological observations—including cosmic microwave background measurements \cite{refId0} and the fifth-year distance measurements of Type Ia supernovae from the Dark Energy Survey \cite{descollaboration2024darkenergysurveycosmology}—we can derive new constraints and insights into the evolution of the cosmological parameters related to the bound dark energy particle $\phi$.

\begin{table*}[ht!]
\caption{\label{tab: Results of some parameters in the BDE, CPL, LCDM for the full dataset} The table shows some key cosmological parameter results from the joint analysis of the DESI DR1 BAO data \cite{desicollaboration2024desi2024vicosmological}, the full CMB dataset \cite{refId0} and the Dark Energy Survey SNeIa compilation from the fifth year \cite{descollaboration2024darkenergysurveycosmology} for bound dark energy, standard $\Lambda$CDM, and $w_0w_a$CDM cosmological models. The results presented are the best fit, the marginalized means, and $68\%$ intervals in each case. TThe reduced $\chi^2$ values for DESI BAO, CMB, and DES-SN5YR for BDE, $\Lambda$CDM, and $w_0w_a$CDM models are shown at the bottom of the table.}
\begin{ruledtabular}
\begin{tabular}{lp{0.4in}lllll}
& \multicolumn{2}{c}{BDE} & \multicolumn{2}{c}{$w_0w_a$CDM} & \multicolumn{2}{c}{$\Lambda$CDM}\\
 Parameters & Best fit & $68\%$ limits & Best fit & $68\%$ limits & Best fit & $68\%$ limits\\ \hline
 $10^{6} a_{c}$     & 2.4862     & 2.4972    $\pm$ 0.0108     &    ---     &          ---             & ---      &  ---  \\
 $\Lambda_{c}$[eV ] & 43.9987    & 43.806   $\pm$ 0.19     &    ---     &          ---             &  ---     & ---    \\
 $w_{0}$            & -0.9300  &-0.9301  $\pm$ 0.0004   & -0.7139    & -0.7240 $\pm$ 0.0712     &  -1      &  -1    \\
 $w_{a}$            & -0.8102  & 0.8085  $\pm$ 0.0053   & -1.1128    & -1.068$^{+0.35}_{-0.30}$ &  ---     & ---    \\
 $H_{0}$            & 67.65    & 67.26   $\pm$ 0.36     & 67.27      & 67.20 $\pm$ 0.65         & 67.86    & 67.79 $\pm$ 0.40\\
 $\Omega_{m}$       & 0.307    & 0.312   $\pm$ 0.005    & 0.316      & 0.316 $\pm$ 0.006        & 0.308    & 0.309 $\pm$ 0.005\\
 $\Omega_{DE}$      & 0.692    & 0.687   $\pm$ 0.005    & 0.683      & 0.683 $\pm$ 0.006        & 0.691    & 0.690 $\pm$ 0.005\\
 $\sigma_{8}$       & 0.805    & 0.810   $\pm$ 0.007    & 0.825      & 0.814 $\pm$ 0.012        & 0.807    & 0.808 $\pm$ 0.007\\
 $100\theta_{MC}$   & 1.04117  & 1.04101 $\pm$ 0.00029  & 1.04094    & 1.04093 $\pm$ 0.00029      & 1.04110  & 104106 $\pm$ 0.00028\\
$D_{A}(z_{*})$      & 13.89    & 13.88   $\pm$ 0.02   &  13.86      & 13.87 $\pm$ 0.02   & 13.89   & 13.89   $\pm$ 0.02 \\
$r_{*}$             & 144.64   & 144.50  $\pm$ 0.21   & 144.38      & 144.44 $\pm$ 0.25  & 144.68  & 144.64 $\pm$ 0.21\\
$z_{*}$             & 1089.81  & 1090.02 $\pm$ 0.21   & 1089.87     & 1089.92 $\pm$ 0.23 & 1089.74 & 1089.76 $\pm$ 0.20 \\
$r_{d}$             & 147.21   & 147.11  $\pm$ 0.23   & 147.03      & 147.10 $\pm$ 0.25  & 147.33  & 147.29 $\pm$ 0.22 \\
$z_{eq}$            & 3371     & 3375    $\pm$ 21     &  3404       & 3401 $\pm$ 26      & 3376    & 3380 $\pm$ 21 \\
\end{tabular}
\begin{tablenotes}
\item $\chi^{2}_{red}(\text{BDE})=1.73 (\text{DESI BAO}) + 0.908 (\text{DES-SN5YR}) + 0.602(\text{CMB})$
\item $\chi^{2}_{red}(\Lambda\text{CDM}) = 2.76(\text{DESI BAO }) + 0.904 (\text{DES-SN5YR}) + 0.599(\text{CMB}) $
\item $\chi^{2}_{red}(w_0w_a\text{CDM}) = 3.00(\text{DESI BAO}) + 0.904 (\text{DES-SN5YR}) + 0.598(\text{CMB}) $
\end{tablenotes}
\end{ruledtabular}
\end{table*}
\begin{figure}[ht]
\includegraphics[width=\linewidth]{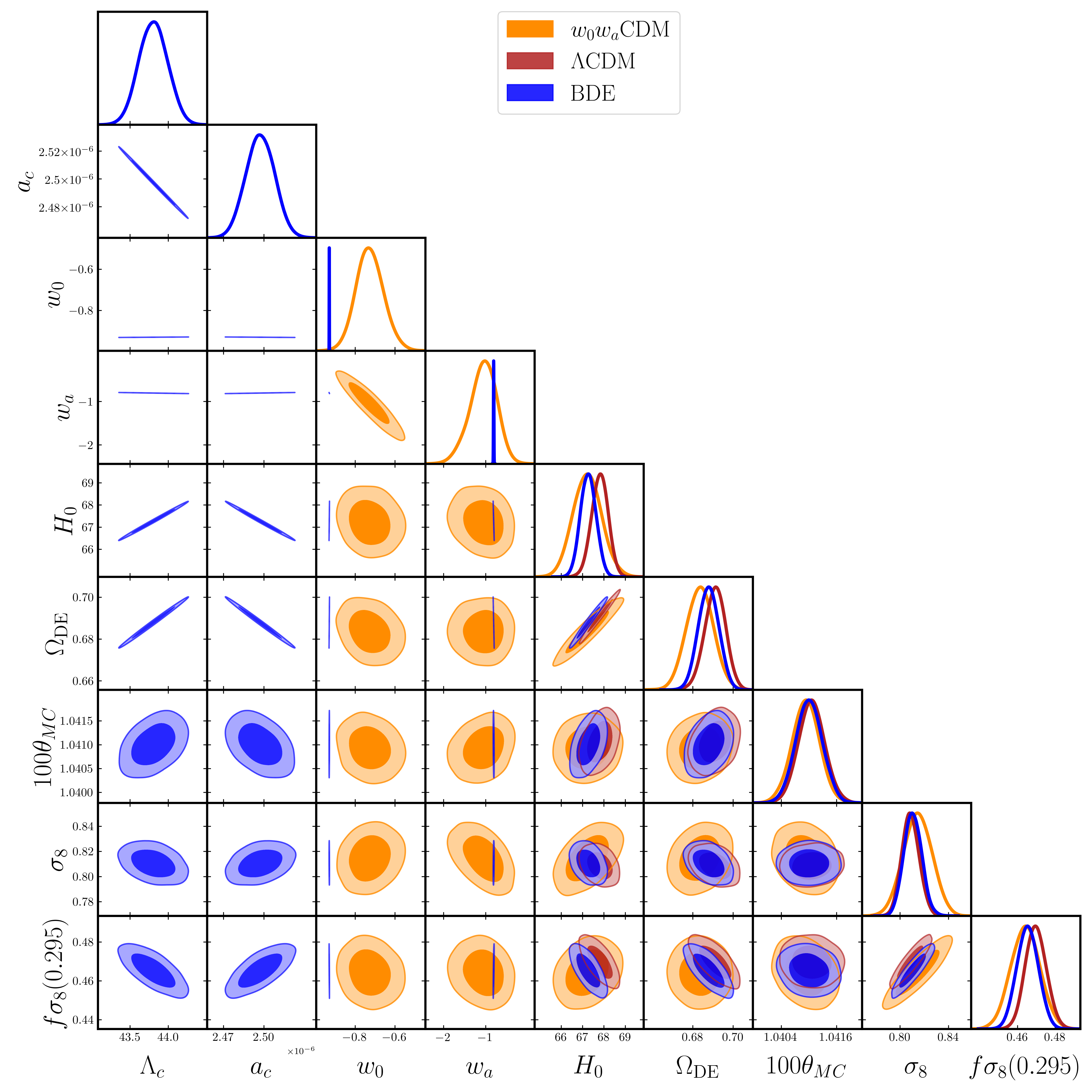}
\caption{Marginalised $68\%$ and $95\%$ posterior constraints on $\Lambda_{c}$, $a_c$, $w_0$, $w_a$, $H_0$, $\Omega_m$, $\Omega_{DE}$, $\sigma_8$ in the BDE, $\Lambda$CDM and
w0waCDM model from DESI DR1 BAO \cite{desicollaboration2024desi2024vicosmological}, CMB\cite{refId0}, and DES-SN5YR datasets \cite{descollaboration2024darkenergysurveycosmology}.}
\label{Fig:2}
\end{figure}

\begin{figure*}[ht]
\includegraphics[width=\textwidth]{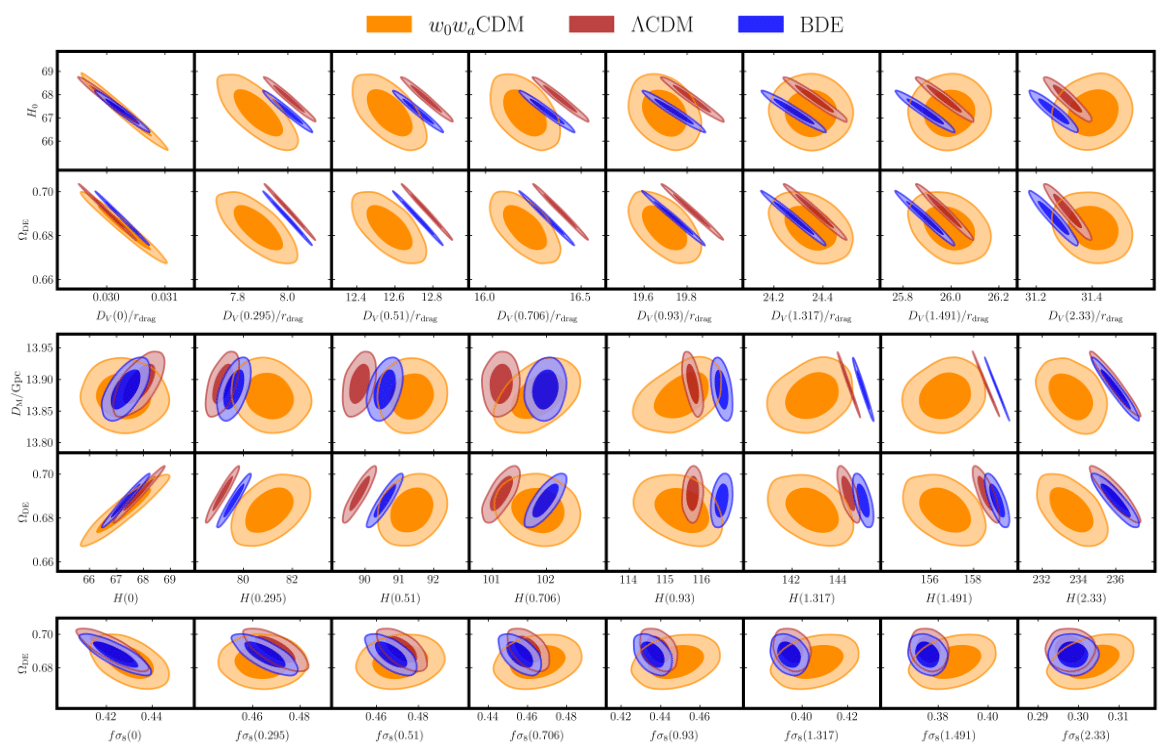}
\caption{\label{Fig:3} Marginalized distributions and $68\%$ and $95\%$ confidence contours of the redshift-distance measurements encoded in $D_{V}(z_{eff})/r_{d}$, $H(z_{eff})$, and $f\sigma_8(z_{eff})$ parameters are presented at the effective redshifts corresponding to the highest significance of BAO detection from DESI. This is in relation to the present-time Hubble parameter, $H_0$, the density parameter of dark energy, $\Omega_{DE}$, and the comoving angular distance $D_M(z^{*})$. The results are shown for three models: bound dark energy (in red), $\Lambda$CDM (in dark red), and $w_0w_a$CDM (in dark orange). These findings are based on the joint analysis from DESI \cite{desicollaboration2024desi2024vicosmological}, CMB \cite{refId0}, and DES-SN5YR \cite{descollaboration2024darkenergysurveycosmology}. }
\end{figure*}

\noindent\textbf{Theory--} The Bound Dark Energy framework suggests that the lightest mesonic field, denoted as $\phi$, is responsible for the late-time acceleration of the universe (i.e. dark energy) and emerged dynamically as the universe expands and cools down specifically occurring at the condensation energy and scale factor denoted by $\Lambda_{c}$, $a_{c}$, respectively. This emergence is driven by the non-perturbative dynamics of a supersymmetric dark gauge group, where the gauge coupling constant exhibits strong interactions at low energy scales \cite{PhysRevLett.121.161303}. As a result, a phase transition occurs, forming light composite states—namely dark mesons. This process bears resemblance to the QCD interactions in the Standard Model, where neutral composite particles (baryons and mesons) are generated from underlying quark constituents. The evolution of this meson field is governed by a non-perturbative inverse-power-law (IPL) scalar potential, $V(\phi)$, which is generated within a strong gauge coupling regime of a supersymmetric gauge group \cite{BURGESS1997181,AFFLECK1984493}. This potential can be mathematically described by the equation \cite{PhysRevD.72.043508,PhysRevLett.121.161303}
\begin{equation}
    V(\phi)=\Lambda_{c}^{4+n}\phi^{-n},
\end{equation}
where the value of the exponent $n$ is determined by the choice of the supersymmetric gauge group $SU(N_c)$ \cite{2003JHEP...01..033D,PhysRevLett.121.161303} and its particle content $N_f$  with  $n=2(N_c-N_f)/(N_c+N_f)=2/3$ with  $N_c=3$, $N_f=6$ for our BDE model. The condensation energy scale $\Lambda_c=\Lambda_{gut} e^{-8\pi^2/(b_0 g _{gut}^2)} = 34^{+16}_{-11}\mathrm{eV}$ with $b_0=3N_c-N_f=3$, and  $g^2_{gut}= 4\pi/(25.83\pm 0.16)$, and unification scale $\Lambda_{gut}=(1.05\pm 0.07)\times 10^{16}$ GeV \cite{Bourilkov_2015}.  This DG framework parallels the established Standard Model of particle physics, encapsulated by $SU_{\text{QCD}}(N_{c}=3) \times SU(N_{c}=2)_{L}\times U_{Y}(N_{c}=1) $, which incorporates three generations of particles and dictates the strong, weak and electromagnetic interactions. We emphasize that the parameters $N_c$ and $N_f$ in the Standard Model specify the theoretical constructs but do not arise from any underlying more fundamental theory, setting an equivalent fundamental level between the standard model and BDE. The energy density of bound dark energy before the condensation point ($a \leq a_{c} $) is relativistic and can be expressed as $\rho_{DE}(a) = 3(a_{c} \Lambda_{c})^{4} a^{-4}$. This equation encompasses two fundamental aspects of the bound dark energy model: the energy scale $ \Lambda_{c} $, which characterizes the intensity of dark interaction, and the scale factor \( a_{c} \), where condensation occurs. For $a<a_c$ the energy density of the DG can be expressed as: $\rho_\mathrm{DG}(a)=\rho_\mathrm{DG}(a_c)(a_c/a)^{4} = 3\Lambda_c^4  (a_c/a)^{4}$, with $\rho_\textrm{DG}(a_c) = 3\Lambda_c^4$.   Therefore, it can be rewritten as $\rho_\mathrm{DG}(a_c)/\rho_r(ac)=3\Lambda_c^3/(\rho_{r0}a_c^{-4})=3(a_c\Lambda_c)^4/\rho_{r0}$, and solving for $a_c\Lambda_c$, we arrive at the constraint equation \cite{PhysRevD.99.103504,PhysRevLett.121.161303} (see supplementary section for detailed derivation):
\begin{equation}\label{eq:bde_acLc_theory}
    \frac{a_c\Lambda_c}{\textrm{eV}} = 1.0939 \times 10^{-4}.
\end{equation}
Eq.(\ref{eq:bde_acLc_theory})  is a meaningful restriction relating the two characteristic quantities of our BDE model, namely, the energy scale $\Lambda_c$ and the scale factor $a_c$ at the condensation epoch. Once the system passes through the condensation phase ($a > a_{c} $) the dark energy density and pressure of the scalar field  exhibit characteristics akin to quintessence as shown in fig.(\ref{Fig:1}), expressed as $\rho_{DE} = \dot{\phi}^{2}/2 + V, \quad P_{DE} = \dot{\phi}^{2}/2 - V$. The dark energy equation of state, given by $w_{DE} = P_{DE}/\rho_{DE} = (\dot{\phi}^{2}/2 - V)/ (\dot{\phi}^{2}/2 + V)$, is time-dependent and influenced by the interplay between the kinetic term $\dot{\phi}^{2}/2 $ and the potential $V(\phi)$. The dynamics of bound dark energy adhere to a Friedmann-Lemaître-Robertson-Walker (FLRW) metric, described by the equation $\ddot{\phi} + 3H\dot{\phi} + \mathrm{d}V/\mathrm{d}\phi = 0$, with the Hubble parameter defined as $H \equiv \dot{a}/{a} = \sqrt{8\pi G \rho_{tot}/3}$. The total energy density of the universe can be summarized as $\rho_{tot}(a)= \rho_{m0} a^{-3} + \rho_{r0} a^{-4} + \rho_{BDE}$, where $\rho_{BDE} = \dot{\phi}^{2}/2 + V(\phi)$. This model provides a concise understanding of origin and dynamics dark energy and its role in cosmic evolution.

\noindent \textbf{Results--} The recent findings from the Dark Energy Spectroscopic Instrument (DESI), particularly those concerning baryon acoustic oscillations (BAO), introduce a transformative perspective on dark energy by suggesting its potential evolution over time. The DESI BAO data favors a dynamical model of dark energy, characterized by a redshift-dependent equation of state parameter $w(z)$. This deviates from the conventional assumption of a cosmological constant, typically represented by $w_0 = -1$ and $ w_a = 0 $. The observed shifts in these parameters reflect a statistical significance ranging from $2.5\sigma$ to $3.9\sigma$, particularly when the data is synergistically analyzed alongside Planck cosmic microwave background (CMB) measurements and supernova observations from the DES-SN5YR program \cite{desicollaboration2024desi2024vicosmological}. In light of the DESI findings, the bound dark energy model reveals that dark energy indeed diverges from a cosmological constant. Figure \ref{Fig:1}\textbf{(a)} shows the $68\%$ and $95\%$ confidence contours for the BDE and $w_0w_a$CDM model. A striking characteristic of this comparison is that the area encompassed by the confidence contours for the BDE model is an astonishing $\sim 10,000$ times smaller than that of the $w_0 w_a$CDM model. This significant reduction in contour area highlights a crucial aspect of the BDE model's robustness due to its null-free parameter nature and theoretical priors, in comparison with $w_0w_a$CDM model. Notice that this parametrization is a picture of equation of state at present time. The observational estimates at $68\%$ C.L. for the parameters that encode the dark energy dynamics are $w_0 = -0.9301 \pm 0.0004 $ and $w_a = -0.8085 \pm 0.0053 $ with respect to the $w_{0}w_{a}CDM$ model yielding $w_{0} = -0.724 \pm 0.0701$ and $w_{a} = -1.06^{+0.31}_{-0.27}$ as reported in \cite{desicollaboration2024desi2024vicosmological}.  Furthermore, the evolution of the equation of state of dark energy in the range of $0<z<3$ and a closer view for BDE, $\Lambda$CDM, and $w_0w_a$CDM models  are visualized in Fig. \ref{Fig:1}\textbf{(b)} and Fig \ref{Fig:1}\textbf{(c)}, respectively, along with their $68\%$ and $95\%$ contour region bands. Figure \ref{Fig:1}\textbf{(d)} offers a wider picture of the evolution of the dark energy equation of state parameter $w(a)$ across the given models as a function of the scale factor $a$.  Before the condensation epoch ($a_{c}$), the equation of state mimics radiation dominance ($w=1/3$) at early times $(a\sim10^{-5})$. At $a_c$, a phase transition occurs as the dark meson field $\phi$ forms non-asymptotically converges to quintessence-like dynamics $w_0=-0.9301$ at present time ($a=1$). From this detailed description, the BDE model shows a sharp, non-monotonic evolving equation of state that remains larger than $w=-1$. On the contrary, $w_0w_a$CDM evolves from values below $w<-1$ (high $z$) to $w>-1$ (low $z$) entering a phantom-like regime which leads to theoretical instabilities associated with negative kinetic energy. Due to the influence of dark energy, Fig. \ref{Fig:1}\textbf{(e)} presents the derived constraints, with shaded regions representing the $68\%$ and $95\%$ confidence intervals for the conformal expansion rate $H(z)$, rescaled by $1 + z$, across the redshift range $0 < z < 3$. The analysis reveals that each of the three examined cosmological models predicts a phase of accelerated expansion, characterized by a negative slope, occurring at redshifts of $z \leq 0.6$. Fig. \ref{Fig:1}\textbf{(f)} presents the matter power spectrum at $z=0$ for the best-fit models of BDE, $\Lambda$CDM and $w_0w_a$CDM. The growth rate of matter overdensities is notably influenced at small scales. Initially, the amplitude of the modes that cross the horizon before $ a_c $ is suppressed compared to $\Lambda$ CDM and $w_0w_a$CDM due to the free streaming of the particles from the dark generator (DG). Subsequently, these modes experience an enhancement caused by the rapid dilution of the bound dark energy (BDE) just after condensation, followed by a slight suppression due to differing dynamics and the contributions of dark energy perturbations in our model. As a result, the matter power spectrum in BDE is enhanced at small scales, with differences from $\Lambda$CDM and $w_0w_a$CDM increasing to $16\%-18\%$ at $k \approx 4.3 \, \mathrm{Mpc}^{-1} $ \cite{Almaraz_2020}. However, these modes are no longer in the linear regime, and we will reserve the analysis of how non-linear dynamics affect this signature for future work. Notably, our derived value for $\Lambda_{c} = 43.806 \pm 0.19$ aligns with the theoretical constraint of $\Lambda_c^{\text{th}} = 34^{+16}_{-11} \, \text{eV} $ within the $68\%$ credible interval, which is based on high-energy physics datasets. Table \ref{tab: Results of some parameters in the BDE, CPL, LCDM for the full dataset} presents the best-fit cosmological parameters and their corresponding constraints at the $68\%$ credible interval. Figure \ref{Fig:2} illustrates that key cosmological parameters, which characterize the expansion rate and energy density components, are in agreement within $1\sigma$ across all three models analyzed, based on the joint interpretation of the provided astrophysical data \cite{descollaboration2024darkenergysurveycosmology, desicollaboration2024desi2024vicosmological, refId0}. Figure \ref{Fig:3} illustrates the $68\%$ and $95\%$ marginalized posterior constraints for the quantities $D_V(z_{eff})/r_d$, $H(z_{eff})$, and $f\sigma_8(z_{eff})$ at the effective redshifts associated with the DESI BAO data, incorporating parameters such as $\Omega_{DE}$, $H_0$, and $D_M$. Notably, the contours of confidence levels exhibit a pronounced splitting behavior as the effective redshift increases. This trend suggests tensions with the standard $\Lambda$CDM model that exceed $2\sigma$, while paradoxically maintaining compatibility with the $w_0w_a$CDM model across most effective redshifts. However, this is not the case for $f\sigma_8$ with $\Omega_{DE}$ where the three models agree between each other at $1\sigma$ level.\\
In assessing the fit quality for recent Baryon Acoustic Oscillation (BAO) measurements from the DESI experiment, a comparison is made between models including $\Lambda\mathrm{CDM}$, $w_0w_a$CDM, and the BDE model. The $\Lambda\mathrm{CDM}$ framework incorporates six free parameters, while the $w_0w_a$CDM variant adds two additional parameters, $ w_0 $ and $ w_a $, resulting in a total of eight free parameters. Conversely, the BDE model encapsulates five free parameters, which denote the current quantity of dark energy, treating its dynamics and magnitude as derived aspects. The DESI BAO dataset comprises twelve observational data points. An analysis of the $\Lambda\mathrm{CDM}$model yields a reduced $(\chi^2)^{\Lambda\mathrm{CDM}}_{\mathrm{BAO}} = 16.56/(12-6) = 2.76$. For the $w_0w_a$CDM model, the calculation gives $(\chi^2)^{w_0w_a\mathrm{CDM}}_{\mathrm{BAO}} = 12.01/(12-8) = 3$. The reduced $\chi^2$ for the BDE model is computed as $(\chi^2)^{\mathrm{BDE}}_{\mathrm{BAO}} = 12.11/(12-5) = 1.73$, with its five free parameters. In terms of the fit of the five-year supernova survey data from DES, the values of $\chi^{2}_{red}$ are mostly equivalent for all three models: BDE ($\chi^{2}_{red}=0.908$), $\Lambda$CDM ($\chi^{2}_{red}=0.904$), and $w_0w_a$CDM ($\chi^{2}_{red}=0.904$). The same goes for CMB: BDE ($\chi^{2}_{red}=0.602$), $\Lambda$CDM ($\chi^{2}_{red}=0.599$), and $w_0w_a$CDM ($\chi^{2}_{red}=0.598$).  In particular, the BDE model results in a $42.35\%$ reduction in the reduced $\chi^2_{\mathrm{BAO}}$ compared to the $w_0w_a$CDM model and a $37.29\%$ decrease relative to the $\Lambda\mathrm{CDM}$ model. The marked reductions in the chi-squared statistic reveal a significantly enhanced alignment of the model with observational data, emphasizing the potential of bound dark energy as a robust and effective cosmological framework. This improvement not only highlights its relevance in current theoretical discourse, but also suggests profound implications for the pursuit of new physics that transcends established paradigms, inviting deeper exploration into the fundamental nature of the universe.

\noindent \textbf{Conclusions--} The Bound Dark Energy model introduces a compelling theoretical framework that bridges particle physics and cosmology, proposing that dark energy arises dynamically from the lightest meson field, $\phi$, formed through non-perturbative interaction in a dark gauge group. This model, rooted in an extension of the Standard Model, eliminates the need for ad hoc assumptions about the cosmological constant $\Lambda$ or the $w_0w_a$ parametrization,
provides a natural explanation for the late-time cosmic acceleration. \\
Crucially, BDE predicts a state equation that changes over time $w(z)$, transitioning from radiation-like behavior ($w=1/3$)  prior to the condensation phase transition at $a_c$ to a  fast diluting epoch with with $w=1$  and $\rho \propto 1/a^6$, to later reemerged with $w\simeq -1$ and finally end up at its present value ($w= 0.9301 \pm 0.0004)$ at late times.
This evolution aligns with DESI's preference for $w>-1$ while maintaining consistency with CMB and supernova data. The time-dependent equation of state $w(z)$ deviates significantly from the cosmological constant ($w_0=-0.9301 \pm 0.0004$, $w_a=-0.8085 \pm 0.0053$), which favors a dynamic dark energy scenario. In particular, BDE achieves these results with three parameters less than $w_0w_a$CDM, enhancing its predictive power and simplicity. Using a state-of-the-art cosmological dataset-including DESI DR1 BAO, Planck CMB, and DES-SN5YR supernovae-we demonstrate that BDE offers a statistically superior fit to observations compared to both the $\Lambda$CDM and $w_0w_a$CDM models. Specifically, the BDE model achieves a 37$\%$-42$\%$ reduction in the reduced $\chi^{2}$ for BAO data, while maintaining consistency with other datasets. Key parameters of the model, such as the condensation energy scale $\Lambda_{c}=43.806\pm0.19$ eV and the condensation epoch $a_{c}=2.4972\pm0.0108 \times10^{-6}$, align with theoretical predictions derived from high-energy physics. The model also predicts distinctive patterns imprints on the early Universe, such as suppressed growth of matter perturbations at small scaleable decrease in contour area underscores vital attributess and transient dark radiation prior to the condensation epoch. While these effects remain consistent with current observational bounds, future high-precision measurements of the CMB, large-scale structure, and primordial element abundances will further test the BDE framework. By addressing the cosmological constant problem through a particle-physics-motivated mechanism and improving the fits to observational data, the BDE model emerges as a viable alternative to $\Lambda$CDM and $w_0w_a$CDM models. Its success underscores the potential of composite dark sector particles to resolve fundamental cosmological puzzles and highlights the importance of interdisciplinary approaches in unraveling the nature of dark energy.

%
\begin{acknowledgments}
\noindent We acknowledge financial support from: PAPIIT-DGAPA-UNAM 101124. Jose Agustin Lozano Torres acknowledges the support of a doctoral scholarship from CONAHCYT.
\end{acknowledgments}
\section*{Author contributions}
\noindent The Bound Dark Energy (BDE) model was originally proposed by Axel de la Macorra. The analysis of these results, interpretations, and constraints were discussed by Axel de la Macorra and Jose Agustin Lozano Torres. The manuscript was written by Jose Agustin Lozano Torres and Axel de la Macorra. 
\section*{Author information}
\noindent Reprints and permission information are available at www.nature.com/reprints. Correspondence and requests for materials should be addressed to Jose Agustin Lozano Torres \url{jalozanotorres@gmail.com} or Axel de la Macorra \url{macorra@fisica.unam.mx}.
\section*{Financial interests}
\noindent The authors declare no competing financial interests.
\section*{Data availability}
\noindent The datasets analyzed in this study can be accessed at the following URLs: for the DESI data \url{https://data.desi.lbl.gov/doc/releases/dr1/}, for the DES-SN5YR data \url{https://github.com/des-science/DES-SN5YR}, and for the CMB data \url{https://pla.esac.esa.int/#cosmology}.
\section*{Code availability}
\noindent The analysis codes will be made available on reasonable request.


\nocite{*}
%

\clearpage

\section{Supplementary information}

\subsection*{Methodology} 

The cosmological data employed to investigate the implications and constraints of various cosmological parameters in three distinct models: the bound dark energy model, the standard model of cosmology, and the $w_0w_a$ CDM model—were meticulously curated from a variety of state-of-the-art astrophysical experiments. This encompassed the analysis of baryon acoustic oscillations, using data from the first year of data release of the Dark Energy Spectroscopic Instrument (DESI) survey \cite{desicollaboration2024desi2024vicosmological}. In addition to this, we systematically evaluated the power spectra of temperature anisotropies (TT) and polarization (EE) auto-spectra, as well as their cross-spectra (TE), which were integrated through advanced likelihood analyses implemented in software packages such as \texttt{simall}, \texttt{Commander} (for multipoles with $\ell < 30$), and \texttt{plik} (for $\ell \ge 30$), all derived from the seminal \textit{Planck} release \cite{refId0}. Moreover, our analysis was further enriched by a comprehensive compilation of 1,635 confirmed Type Ia supernovae, sourced from the Dark Energy Survey (DES) \cite{descollaboration2024darkenergysurveycosmology}. This dataset, part of their Year 5 data release, encompasses a remarkable spectrum of distances—from relatively nearby celestial objects to those positioned at the extreme fringes of the observable universe, characterized by redshift values ranging from $0.1$ to $1.3$. To meticulously analyze this extensive and diverse dataset, we employed an advanced Markov Chain Monte Carlo (MCMC) analysis technique \cite{Lewis_2002}, using modified versions of the well-established cosmological codes CAMB \cite{Lewis_2000} and CosmoMC \cite{Lewis_2002}. The Metropolis-Hastings MCMC sampler was leveraged to run four independent chains in parallel for each specific dataset and model combination, initiating the process with proposal covariance matrices derived from preliminary simulation runs. The chains were persistently executed until they satisfied the stringent Gelman-Rubin criterion of $R - 1 < 0.01$, thus ensuring the robustness and statistical reliability of the results obtained. To derive the constraints presented in our findings, we skillfully harnessed the capabilities of the Python package \textsc{getdist} \cite{2019arXiv191013970L}. Furthermore, we meticulously calculated the best-fit values for each cosmological model utilizing the \texttt{Powell's 2009 BOBYQA} bounded minimization routine, as integrated into \texttt{CosmoMC}. This process involved executing four independent minimizations from various random starting points to ensure convergence and significantly enhance the reliability and integrity of our findings. We aim to provide a comprehensive and nuanced understanding of the underlying cosmological frameworks through these methodologies.

 \subsection*{ Theoretical Framework of BDE}
The Standard Model (SM) of particle physics stands as a monumental framework that intricately describes the interactions among the universe's fundamental particles, including photons, electrons, quarks, and protons. These interactions are governed by three distinct gauge groups: SU(3), SU(2), and U(1), which correspond to the strong nuclear force, weak interactions, and electromagnetic forces, respectively. This model is renowned for being one of the most accurate theories in the realm of physics, providing profound insights into the behavior of matter at the most fundamental level. However, despite its remarkable success, the SM has notable shortcomings. It does not incorporate the mysteries of dark energy (DE) and dark matter (DM), which together account for a staggering $96\%$ of the universe's total matter-energy content—$68\%$ attributed to dark energy and $28\%$ to dark matter. These enigmatic components elude direct detection, yet they play a crucial role in the cosmic landscape. In our investigation, we delve into the Bound Dark Energy (BDE) model, a theoretical framework that seeks to shed light on these elusive phenomena. Initially proposed in 1995 and further explored in 2018 \cite{PhysRevLett.121.161303}, the BDE model now stands on the brink of transformative understanding. With the extraordinary data and measurements provided by DESI, we can now discern the dynamics of dark energy with unprecedented clarity, allowing us to distinguish between different cosmological scenarios and deepen our comprehension of the universe's fundamental nature\\   
We propose a dark gauge group, whereby its gauge coupling constant becomes strong at low energies. This leads to a phase transition that generates light composite states, such as dark mesons, akin to the strong (QCD) interaction in which neutral QCD particles (baryons and mesons) arise from fundamental quarks.  The Bound Dark Energy (BDE) model features a supersymmetric dark gauge group denoted "DG," characterized by $SU(N_c=3)$ and $N_f=6$ particles in the fundamental representation. The parameters $N_c$ and $N_f$ are fundamental to defining our model, which holds a status equivalent to that of Standard Model interactions. Consequently, BDE shares a foundational significance with the well-established Standard Model of particle physics, represented as $SU_\textrm{QCD}(N_c=3) \times SU(N_c=2)_L \times U_Y(N_c=1)$, which encompasses three families that describe the strong, weak, and electroweak interactions.\\ 
As the universe expands, it cools, resulting in a decrease in both energy densities and temperature. Concurrently, Baryon Dark Energy (BDE) and Quantum Chromodynamics (QCD) interactions intensify, leading to the formation of neutral particles such as neutrons and pions in QCD, as well as bound dark particles within the BDE framework. From this point onward, we will refer to the BDE particle as the lightest meson field, denoted by $\phi$, which arises from the non-perturbative dynamics of the BDE group. This evolution is characterized by an inverse power law potential (IPL), expressed as $V(\phi)=\Lambda_c^{4+n}\phi^{-n}$, with exponent $n=2[1+2/(N_c-N_f)]=2/3$. This formulation is derived from the non-perturbative Affleck-Dine-Seiberg superpotential (ADS) \cite{AFFLECK1984493}, where the value of $n$ is determined by the number of colors ($N_c=3$) and flavors ($N_f=6$) in the BDE gauge group. We find it appealing to assume the unification of gauge couplings among the standard model (SM) gauge groups, which correspond to strong, weak and electromagnetic interactions, and the BDE gauge group, on the unification scale $\Lambda_{gut}$. In the Minimal Supersymmetric Standard Model (MSSM), the unification of the coupling constants occurs at the scale $\Lambda_{gut} = (1.05 \pm 0.07) \times 10^{16} \mathrm{GeV}$, with the unified coupling constant given by $g_{gut}^2 = 4\pi/(25.83 \pm 0.16)$ \cite{Bourilkov_2015}. Using these values of $\Lambda_{gut}$ and $ g^2_{gut}$ in the equation of the one-loop renormalization group

\begin{equation}\label{eq:bde_coupling}
g^{-2}(\Lambda) = g_{gut}^{-2} + \frac{b_0}{8\pi^2} \ln \Bigg(\frac{\Lambda}{\Lambda_{gut}}\Bigg), 
\end{equation}

\noindent we obtain a condensation scale \cite{PhysRevLett.121.161303}:

\begin{equation}\label{eq:bde_Lc_theory}
 \Lambda_c=\Lambda_{gut} e^{-8\pi^2/(b_0 g _{gut}^2)} = 34^{+16}_{-11}\mathrm{eV},
\end{equation}

\noindent where $g_{gut} \equiv g(\Lambda_{gut})$ corresponds to the gauge coupling value on the unification scale $\Lambda_{gut} $ and $b_0$ counts the number of elementary particles charged in the gauge group. In our BDE model we have $b_0=3N_c-N_f=3$, with $N_c=3,N_f=6$. The significant exponential suppression of the condensation scale $\Lambda_c$ relative to the grand unification scale $\Lambda_{gut}$ elucidates the reason behind the markedly lower value of $\Lambda_c$. This phenomenon indicates that the phase transition associated with the dynamics of dark energy transpires at a later epoch, facilitating the emergence of Dark Energy in a more contemporary timeframe. Importantly, the condensation scale $\Lambda_c$ emerges as a robust prediction within our BDE framework. Nevertheless, the uncertainties associated with equation (\ref{eq:bde_Lc_theory}) predominantly arise from the inaccurate assessment of $\Lambda_{gut}$ and $g_{gut}^2$, which are closely linked to fluctuations in the QCD gauge coupling constant.

\subsubsection{Initial Conditions  and Gauge Coupling Unification}

At high energy levels, all the particles in the DG and the MSSM become relativistic, allowing the energy density to be represented in terms of the relativistic degrees of freedom ($g$) and the temperature ($T$). For the MSSM, the following expression holds: $\rho_\mathrm{SM}(a)=\frac{\pi^2}{30} g_\mathrm{SM} T_\mathrm{SM}^4(a)$, while the unification of gauge couplings suggests that $T_\mathrm{DG}(a_{gut})=T_\mathrm{SM}(a_{gut})$ at $\Lambda_{gut}$, resulting in the ratio:

\begin{equation}\label{eq:bde_rhoDGrhoSM_gut}
\frac{\rho_\mathrm{DG}(a_{gut})}{\rho_\mathrm{SM}(a_{gut})} = \frac{g_\mathrm{DG}^{gut}}{g_\mathrm{SM}^{gut}}=0.426
\end{equation}

\noindent where $g_\mathrm{SM}^{gut}=228.75$ for the MSSM, and $g_\mathrm{DG}^{gut}=(1+\frac{7}{8})[2(N_c^2-1)+2N_cN_f]=97.5$ \cite{PhysRevD.99.103504}. This indicates that the DG contributes to the $\Omega_\textrm{DG}(a_{gut})=\rho_\textrm{DG}/(\rho_\textrm{SM}+\rho_\textrm{DG})=0.3$ of the universe's overall cosmic composition at that epoch. Below the unification scale, the DG particles only interact with the SM particles through gravitational forces, resulting in the loss of thermal equilibrium between these two categories. Since the DG particles are massless above the energy scale $\Lambda_c$, the number of relativistic degrees of freedom $g_\mathrm{DG}$ remains unchanged until the condensation epoch at $a_c$, whereas the relativistic degrees of freedom $g_\mathrm{SM}$ of the SM decrease as the universe expands and cools. We can apply entropy conservation $S_x=\frac{2\pi^2}{45}g_xT_x^3a^3$ for $x=\mathrm{DG, SM}$ to establish a relationship between their temperatures at various times. It is useful to reference the temperature of the neutrinos $T_\nu$ for $T_\mathrm{SM}$ since they remain in thermal equilibrium with photons for $T > 1 \textrm{ MeV}$, and after neutrino decoupling, electrons and positrons annihilate, slightly increasing the temperature of the photon ($\gamma$) bath above that of the neutrinos, $T_\gamma = \left( \frac{11}{4} \right)^{1/3}T_\nu$. For $a\leq a_c$, entropy conservation results in 

\begin{equation}\label{eq:bde_TDGTSM_grl}
\frac{T_\mathrm{DG}(a)}{T_\mathrm{\nu}(a)}=\left( \frac{g_\mathrm{SM} (a)} {g_\mathrm{SM}^{gut} } \right)^{1/3}
\end{equation}

\noindent and we can use  Eq. (\ref{eq:bde_TDGTSM_grl}) to determine the temperature ratio $T_\mathrm{DG}/T_\mathrm{\nu}(a)$ at different values of $a<a_c$. The introduction of the DG contributes with extra radiation in the early universe using Eq. (\ref{eq:bde_TDGTSM_grl}), the value of $N_{ext}$ in is 

\begin{equation}\label{eq:bde_Next}
N_{ext}=\frac{4}{7}g_\textrm{DG}^{gut} \left(\frac{g_\textrm{SM}^{\nu dec}}{g_\textrm{SM}^{gut}}\right)^{4/3}= 0.945, 
\end{equation}

\noindent valid for $a_{\nu_{dec}}\leq a < a_c$.  At the phase transition at $\Lambda_c$ (i.e. at $a_c$) the elementary particles of the DG form neutral composite states (i.e. the BDE meson particle is formed) and we no longer have extra relativistic particles, so $N_{ext}=0$ for $a\geq a_c$.
By the time when the condensation occurs at $a_c$, only the photons and neutrinos remain relativistic. The energy density of this standard radiation is given by  $\rho_r=\frac{\pi^2}{30}g_rT_\gamma^4$, where $g_r=2+\frac{7}{8}(2)N_\nu \left(T_\nu/T_\gamma\right)^4=3.383$ and $N_\nu=3.046$ account for neutrino decoupling effects Ref[]. Combining this result with $\rho_\mathrm{DG}=(\pi^2/30)g_\mathrm{DG}T_\mathrm{DG}^4$ and Eq. (\ref{eq:bde_TDGTSM_grl}), we find the ratio of the energy density of the DG to the standard radiation at $a_c$:

\begin{equation}\label{eq:bde_rhoDGrhor_ac}
\frac{\rho_\mathrm{DG}(a_c)}{\rho_{r}(a_c)}=\frac{g_\mathrm{DG}^{gut}}{g_r}\left( \frac{4}{11}\frac{g_\mathrm{SM}^{\nu dec}}{g_\mathrm{SM}^{gut}} \right)^{4/3}=0.1268.
\end{equation}

\noindent Since the matter content of the universe is still negligible at that time (i.e. $a\sim 10^{-6}, z\sim 10^6$), this implies that the DG amounts to $\Omega_\textrm{DG}(a_c)=\rho_\textrm{DG}/(\rho_r+\rho_\textrm{DG})=0.113$ of the cosmic budget. For $a<a_c$ the energy density of the DG can be expressed as:

\begin{equation}\label{eq:bde_rho_ac}
\rho_\mathrm{DG}(a)=\rho_\mathrm{DG}(a_c)\Big(\frac{a_c}{a}\Big)^{4}= 3\Lambda_c^4\Big(\frac{a_c}{a}\Big)^{4},
\end{equation}

\noindent with $\rho_\textrm{DG}(a_c) = 3\Lambda_c^4$.  Therefore, Eq. (\ref{eq:bde_rhoDGrhor_ac}) can be rewritten as:

\begin{equation}\label{eq:bde_rhoDGrhor_ac_2}
 \frac{\rho_\mathrm{DG}(a_c)}{\rho_r(ac)}=\frac{3\Lambda_c^3}{\rho_{r0}a_c^{-4}}=\frac{3(a_c\Lambda_c)^4}{\rho_{r0}},
\end{equation}

\noindent and solving for $a_c\Lambda_c$, we arrive at the constraint equation:

\begin{align}\label{eq:bde_acLc_theory_1}
        \frac{a_c\Lambda_c}{\textrm{eV}} & = \left(\frac{\rho_{r0}}{3\textrm{eV}^4} \frac{g_\textrm{DG}^{gut}}{g_r} \right)^{1/4} \left(\frac{4}{11} \frac{g_\textrm{SM}^{\nu_{dec}}}{ g_\textrm{SM}^{gut}}\right)^{1/3}\nonumber \\& = 1.0939 \times 10^{-4}.
\end{align}

Eq.(\ref{eq:bde_acLc_theory_1}) is a meaningful restriction that relates the two characteristic quantities of our BDE model, namely, the energy scale $\Lambda_c$ and the scale factor $a_c$ in the condensation epoch.
\subsubsection{Further cosmological implications}

\begin{figure*}[ht]
\includegraphics[width=1.08\textwidth]{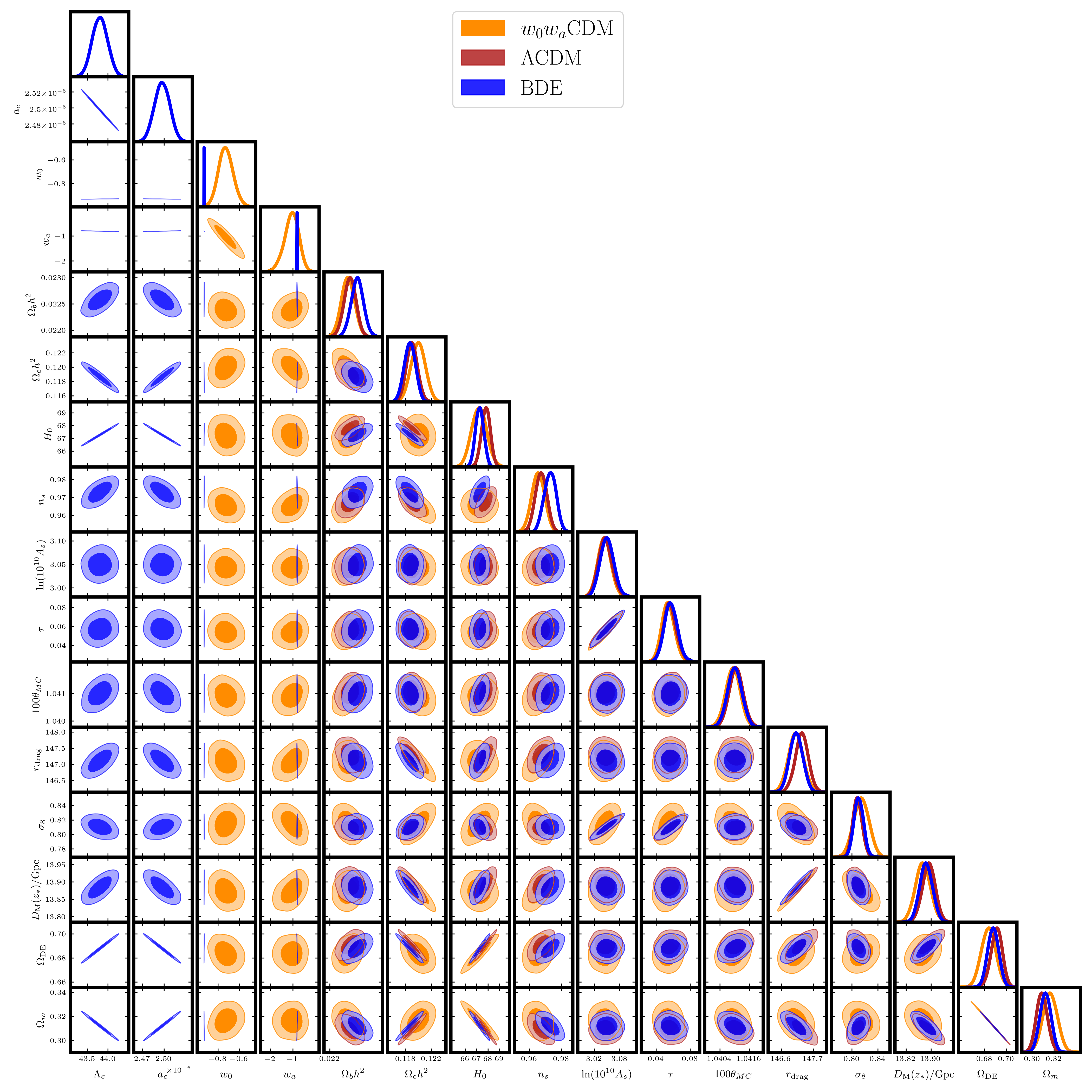}
\caption{\label{Fig:4} Marginalized distributions and $68\%$ and $95\%$ confidence contours of various cosmological parameters for the BDE (red), $\Lambda$CDM (green), and $w_0w_a$CDM models from DESI DR1 BAO \cite{desicollaboration2024desi2024vicosmological}, CMB\cite{refId0}, and DES-SN5YR \cite{descollaboration2024darkenergysurveycosmology}. }
\end{figure*}

\begin{figure*}[ht]
\includegraphics[width=\textwidth]{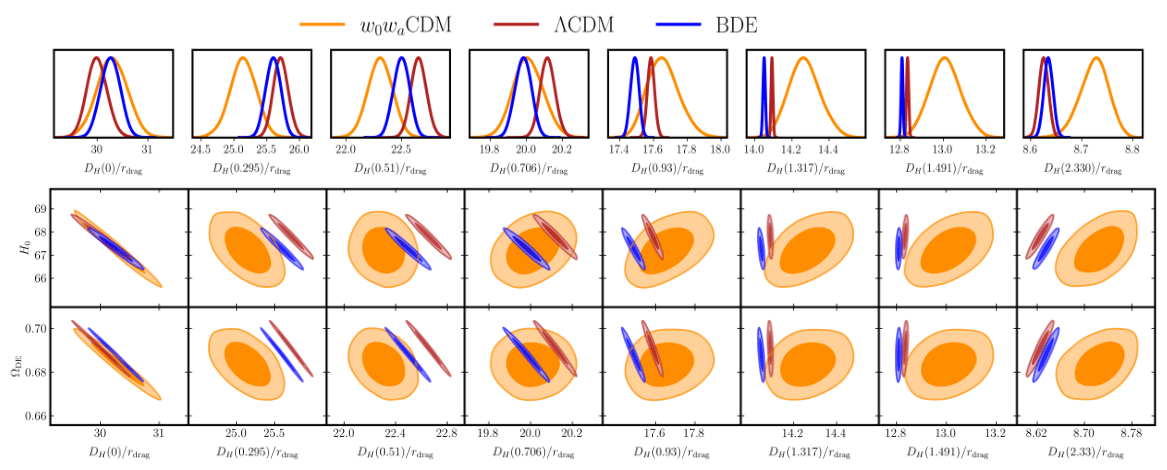}
\caption{\label{Fig:7} Marginalized distributions and $68\%$ and $95\%$ confidence contours of the redshift-distance measurements in terms of $D_{M}/r_{d}$ parameter are presented at the effective redshifts corresponding to the highest significance of BAO detection from DESI. This is in relation to the present-time Hubble parameter, $H_0$, and the density parameter of dark energy, $\Omega_{DE}$. The results are shown for three models: bound dark energy (in red), $\Lambda$CDM (in dark red), and $w_0w_a$CDM (in dark orange). These findings are based on the joint analysis from DESI \cite{desicollaboration2024desi2024vicosmological}, CMB \cite{refId0}, and DES-SN5YR \cite{descollaboration2024darkenergysurveycosmology}.  }
\end{figure*}

\begin{figure*}[ht]
\includegraphics[width=\textwidth]{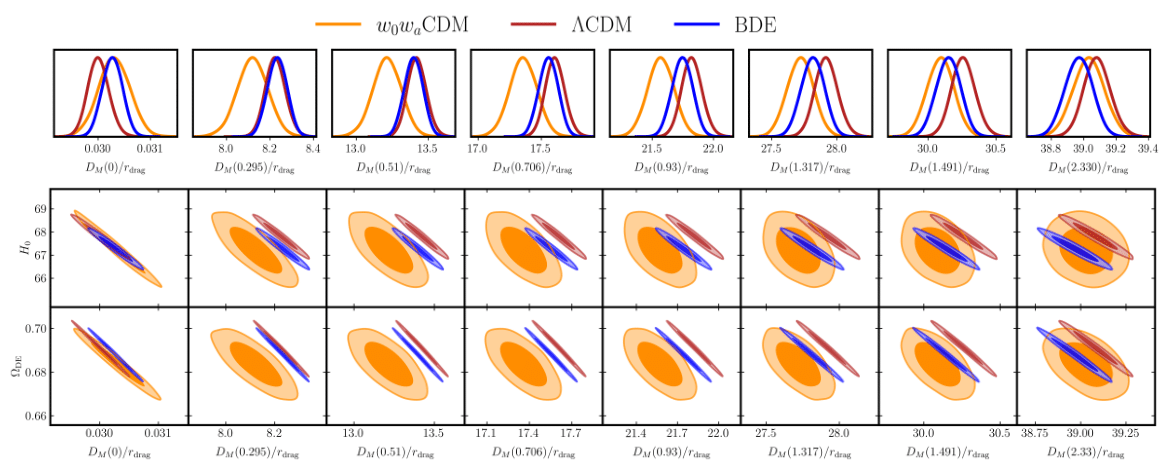}
\caption{\label{Fig:8} Marginalized distributions and $68\%$ and $95\%$ confidence contours of the redshift-distance measurements in terms of $D_{H}/r_{d}$ parameter are presented at the effective redshifts corresponding to the highest significance of BAO detection from DESI. This is in relation to the present-time Hubble parameter, $H_0$, and the density parameter of dark energy, $\Omega_{DE}$. The results are shown for three models: bound dark energy (in red), $\Lambda$CDM (in dark red), and $w_0w_a$CDM (in dark orange). These findings are based on the joint analysis from DESI \cite{desicollaboration2024desi2024vicosmological}, CMB \cite{refId0}, and DES-SN5YR \cite{descollaboration2024darkenergysurveycosmology}.  }
\end{figure*}

\begin{figure*}[ht]
\includegraphics[width=0.98\textwidth]{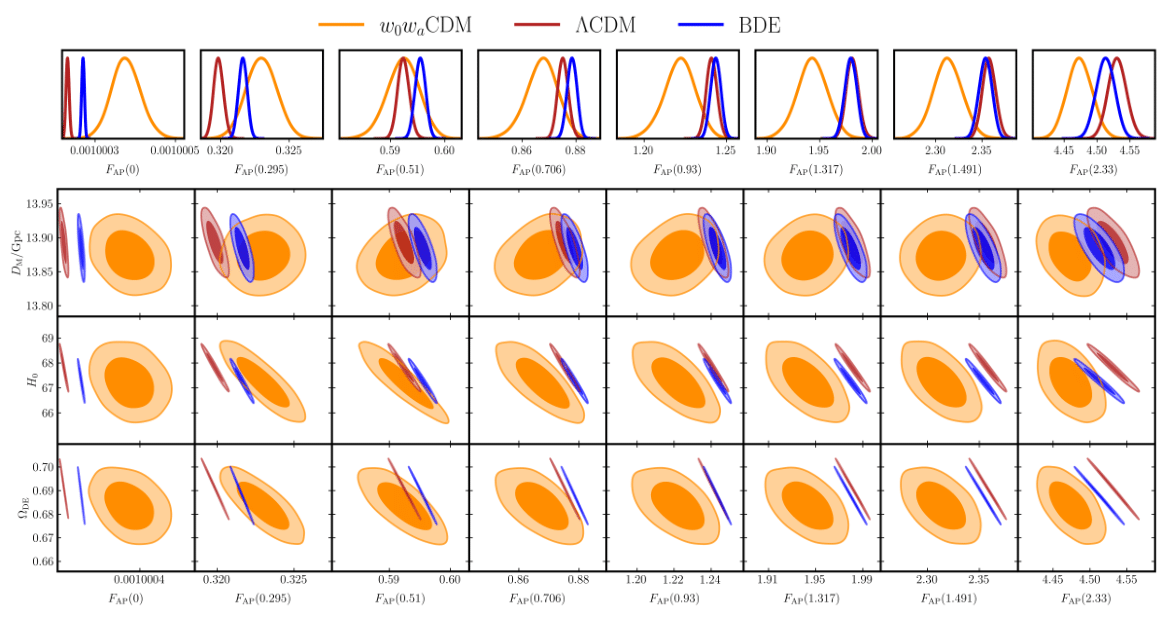}
\caption{\label{Fig:9} Marginalized distributions and $68\%$ and $95\%$ confidence contours of Alcock-Paczinski parameter $F_{AP}(z)$ are presented at the effective redshifts corresponding to the highest significance of BAO detection from DESI. This is in relation to the comoving angular distance $D_M$, the Hubble parameter at present time $H_0$ and the density parameter of dark energy, $\Omega_{DE}$. The results are shown for three models: bound dark energy (in red), $\Lambda$CDM (in dark red), and $w_0w_a$CDM (in dark orange). These findings are based on the joint analysis from DESI \cite{desicollaboration2024desi2024vicosmological}, CMB \cite{refId0}, and DES-SN5YR \cite{descollaboration2024darkenergysurveycosmology}.}
\end{figure*}

\begin{figure*}[ht]
\includegraphics[width=0.8\textwidth]{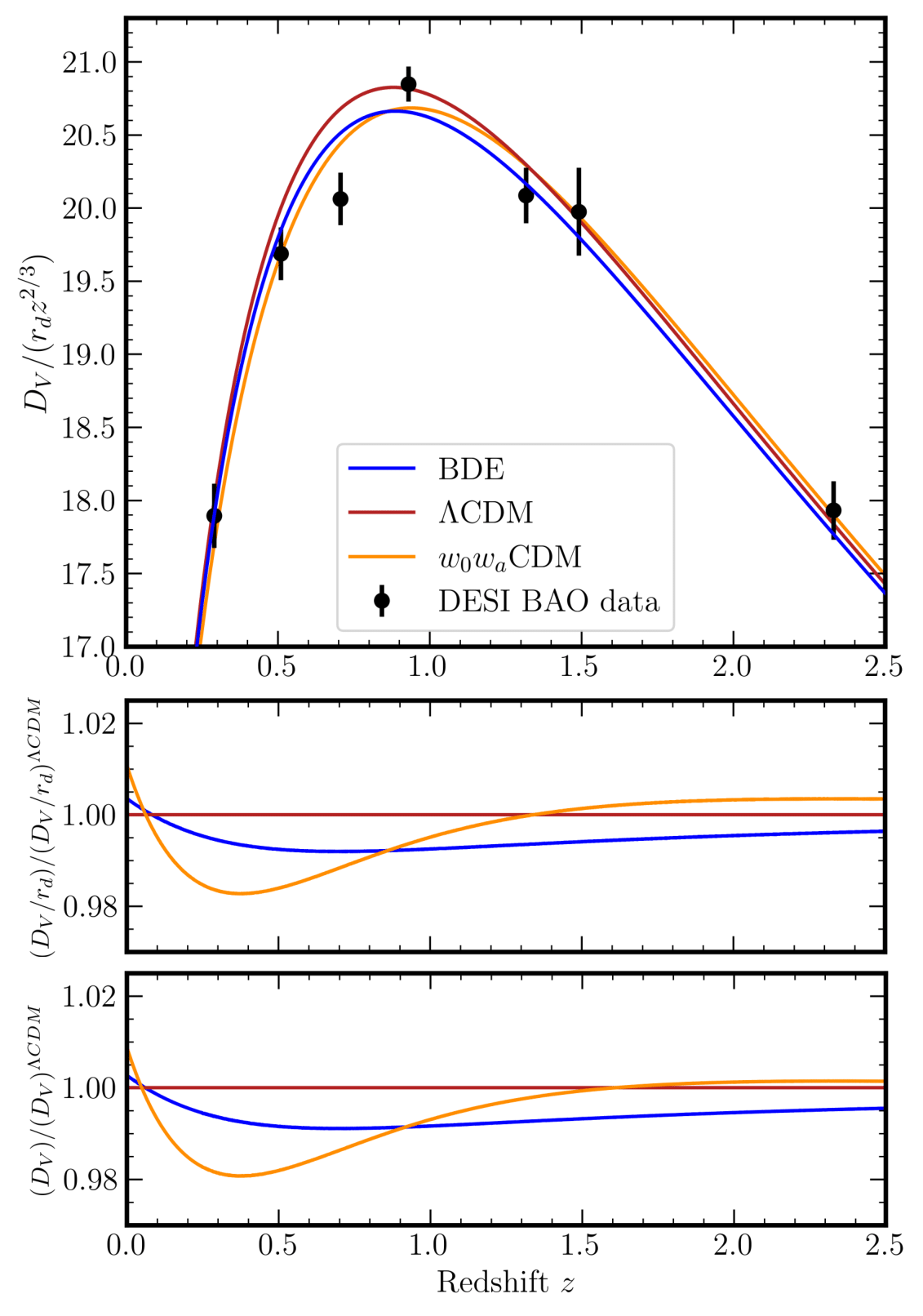}
\caption{\label{Fig:11} The DESI measurements of the Baryon Acoustic Oscillation (BAO) distance scales at various redshifts are parameterized as the ratio of the angle-averaged distance, defined as $D_{V} \equiv (zD^{2}_M D_H)^{1/3}$, to the sound horizon at the baryon drag epoch, $ r_{d} $. These measurements include data from all tracers and redshift bins, as labeled. For clarity, an arbitrary scaling of $ z^{-2/3} $ has been applied. 
The solid lines represent model predictions from different cosmological models: bound dark energy (in blue), the $\Lambda$CDM model (in dark red), and the $w_0w_a$CDM model (in dark orange), based on their best fits. These findings result from a joint analysis involving DESI \cite{desicollaboration2024desi2024vicosmological}, CMB \cite{refId0}, and DES-SN5YR \cite{descollaboration2024darkenergysurveycosmology}. 
The bottom panel illustrates the ratio of the best fits of the bound dark energy and $w_0w_a$CDM models to the best-fit $\Lambda$CDM model. }
\end{figure*}

\begin{figure*}[ht]
\includegraphics[width=0.9\textwidth]{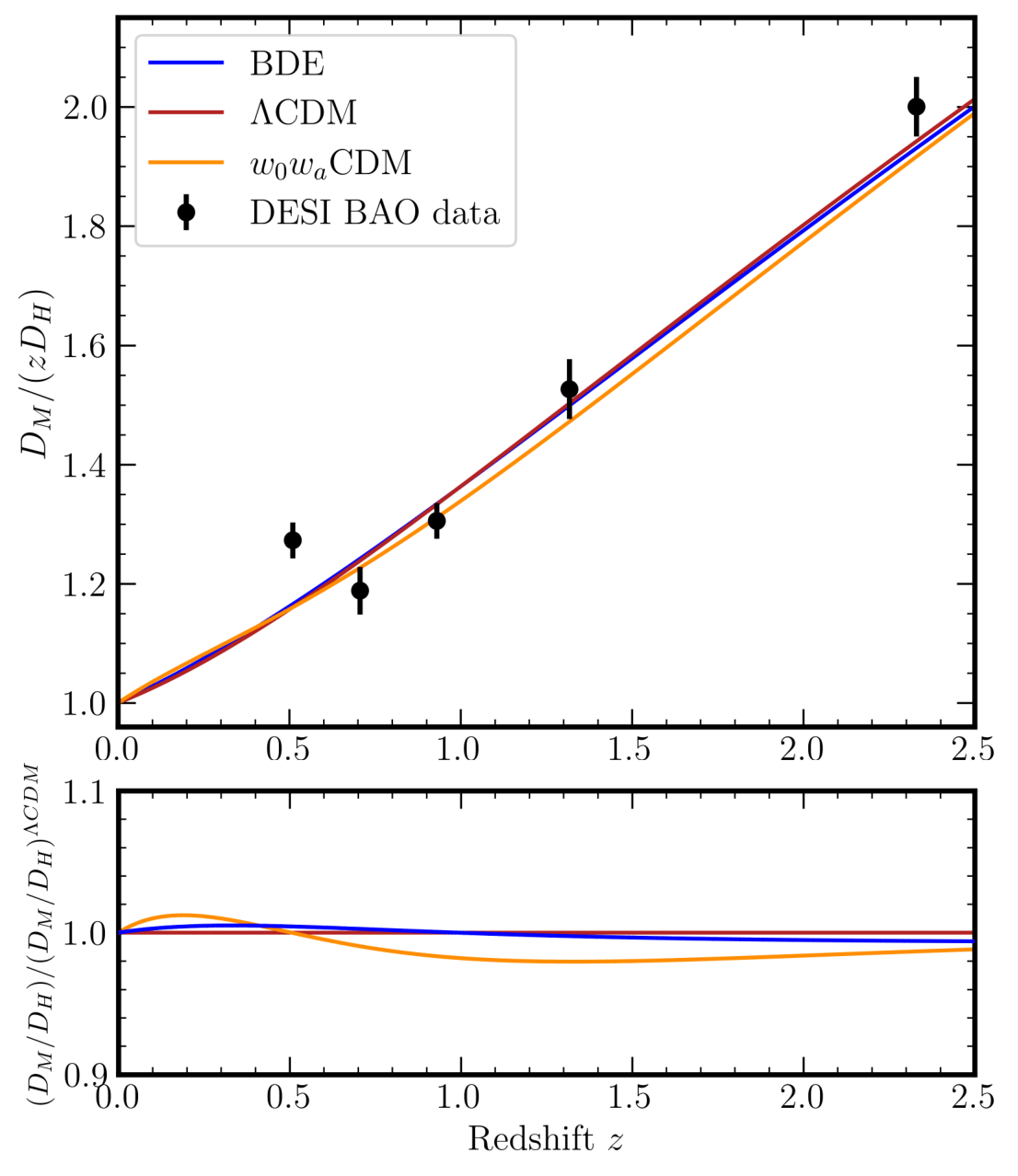}
\caption{\label{Fig:12} The DESI measurements of the Baryon Acoustic Oscillation (BAO) distance scales at various redshifts are parameterized as the ratio of transverse and line-of-sight comoving distances $D_{M}/D_{H}$. These measurements include data from all tracers and redshift bins, as labeled. For clarity, an arbitrary scaling of $ z^{-1} $ has been applied. The solid lines represent model predictions from different cosmological models: bound dark energy (in blue), the $\Lambda$CDM model (in dark red), and the $w_0w_a$CDM model (in dark orange), based on their best fits. These findings result from a joint analysis involving DESI \cite{desicollaboration2024desi2024vicosmological}, CMB \cite{refId0}, and DES-SN5YR \cite{descollaboration2024darkenergysurveycosmology}. 
The bottom panel illustrates the ratio of the best fits of the bound dark energy and $w_0w_a$CDM models to the best-fit $\Lambda$CDM model.}
\end{figure*}

 The discrepancies among various key cosmological parameters are discussed further in this supplementary section, as illustrated in Figure \ref{Fig:4}. This figure presents the marginalized $68\%$ and $95\%$ posterior constraints on six base $\Lambda$CDM parameters, eight parameters from the $w_0w_a$CDM model, BDE parameters, and derived parameters: $\Lambda_{c}$, $a_{c}$, $w_0$, $w_a$, $\Omega_b h^2$, $\Omega_c h^2$, $H_0$, $n_s$, $\tau$, $100\theta_{MC}$, $\ln(10^{10}A_s)$, $H_0$, $r_d$, $\sigma_8$, $\Omega_{DE}$, $\Omega_m$, and $D_M(z^*)$. While most cosmological parameters are in agreement at the $1\sigma$ level across the three models, we observe a notable tension with the $w_0$ parameter between the BDE and $w_0w_a$CDM models, which is at the $2\sigma$ level. The immediate consequence of modifying the expansion rate in the later stages of the universe has a profound effect on the cosmological distances investigated through the evolution of the Hubble distance, denoted as $D_{H}(z_{eff})/r_{d}$, and the angular comoving distance, represented as $D_M(z_{eff})/r_d$. This relationship is particularly evident at specific effective redshifts, as vividly illustrated in Figures \ref{Fig:7} and \ref{Fig:8}. In our observations, we note a significant shift in the distributions for the BDE model and the $\Lambda$CDM model, which migrate from the left side to the right as we transition from a higher redshift of $z=2.33$ to a much lower redshift of $z=0.295$. This trend is further captured in the top panels of Figures \ref{Fig:7} and \ref{Fig:8}, which showcase the dynamic evolution of these distances over time. Moreover, by closely examining the joint constraints involving $H_0$ and $\Omega_{DE}$, we observe that the contours begin to diverge. This separation becomes particularly marked as the distributions associated with $D_{H}(z_{eff})/r_{d}$ and $D_M(z_{eff})/r_d$ in the BDE model shift to the left. This phenomenon leads to a notable tension with respect to the $\Lambda$CDM model, exceeding $2\sigma$ in the context of $D_{H}(z_{eff})/r_{d}$ and the $w_0w_a$CDM model. However, intriguingly, this tension does not manifest similarly in the $D_M(z_{eff})/r_d$ comparison with the $w_0w_a$CDM model, suggesting a more nuanced relationship between these cosmological frameworks. The same characteristics are illustrated in Fig. \ref{Fig:9}.\\
 Figures \ref{Fig:11} and \ref{Fig:12} vividly illustrate the latest results from the Dark Energy Spectroscopic Instrument (DESI) concerning Baryon Acoustic Oscillations (BAO). These results are presented in a detailed Hubble diagram that depicts the angle-average quantities: $D_{V}/r_d$, $D_M/r_{d}$, and $D_{H}/r_d$ for all tracers. In Fig. \ref{Fig:11}, the quantity $D_V/r_d$ is showcased, scaled by an arbitrary factor $z^{-2/3}$, plotted as a function of redshift ranging from 0 to 2.5. This span encompasses all the BAO data points at effective redshifts, providing a comprehensive overview of the model predictions represented by the solid lines. These lines correspond to the best-fitting models from BDE, $\Lambda$CDM, and $w_0w_a$CDM, which align closely with the DESI observations. The lower panels further enrich this analysis, illustrating the comparative ratios of the BDE and $w_0w_a$CDM models against the $\Lambda$CDM benchmark. Conversely, Fig. \ref{Fig:12} displays $D_M/D_H$, scaled by $z^{-1}$, alongside model predictions from BDE, $\Lambda$CDM, and $w_0w_a$CDM models. This figure provides a critical comparison, enhancing our understanding of the implications these models have in the context of current cosmological data. \\
 In summary, we observe a marked difference compared to $\Lambda$CDM and $w_0w_a$CDM models; however, our BDE model remains consistent with astrophysical bounds, and more precise measurements are needed to draw further conclusions. Addressing the problem of dark energy is complex and necessitates introducing new physics. Any alternative scenario aimed at replacing the cosmological constant as the source of cosmic acceleration must not only consistently fit observations but also be grounded in solid theoretical foundations that explain the nature of dark energy and its role in driving the universe's expansion at late times. Furthermore, this candidate must predict deviations from the standard $\Lambda$CDM scenario that can be tested by future observations. A reduction in free parameters and an improved fit to cosmological observations compared to $\Lambda$CDM would provide strong evidence regarding the nature of dark energy. The dark energy model we present here meets all these criteria: BDE has a sound derivation as a natural extension of the Standard Model of particle physics, features no free parameters, and achieves an excellent fit with current cosmological data, particularly enhancing the likelihood of the Baryon Acoustic Oscillation (BAO) distance measurements. With the advent of precision cosmology and the ongoing quest for a new paradigm in the dark sector of the universe, our BDE model presents a compelling framework for exploration in the coming years.

\end{document}